\documentclass[12pt]{iopart}

\usepackage{braket}
\usepackage{graphicx}
	\newcommand{\un}[1]{\mathrm{\:#1}}
\begin{document}
	
	\title{Impact of Non-Unitary Spin Squeezing on Atomic Clock Performance}
	\date{\today}
	\author{B Braverman\footnote{bbraverm@mit.edu; Present address: Department of Physics and Max Planck Centre for Extreme and Quantum Photonics, University of Ottawa, 25 Templeton Street, Ottawa, Ontario K1N 6N5, Canada}, 
			A Kawasaki\footnote{Present address: W. W. Hansen Experimental Physics Laboratory and Department of Physics, Stanford University, Stanford, California 94305, USA} and 
			V Vuleti\'{c}}
	
	\address{Department of Physics, MIT-Harvard Center for Ultracold Atoms and Research Laboratory of Electronics, Massachusetts Institute of Technology, Cambridge, Massachusetts 02139, USA}
	\ead{vuletic@mit.edu}
%	\author{Boris Braverman}
%	\email{bbraverm@mit.edu}
%	\altaffiliation[Current address: ]{Department of Physics and Max Planck Centre for Extreme and Quantum Photonics, University of Ottawa, 25 Templeton Street, Ottawa, Ontario K1N 6N5, Canada}
%	\author{Akio Kawasaki}
%	\altaffiliation[Current address: ]{W. W. Hansen Experimental Physics Laboratory and Department of
%		Physics, Stanford University, Stanford, California 94305, USA}
%	\author{Vladan Vuleti\'{c}}
%	\email{vuletic@mit.edu}
%	\affiliation{Department of Physics, MIT-Harvard Center for Ultracold Atoms and Research Laboratory of Electronics, Massachusetts Institute of Technology, Cambridge, Massachusetts 02139, USA}

	\begin{abstract}
		Spin squeezing is a form of entanglement that can improve the stability of quantum sensors operating with multiple particles, by inducing inter-particle correlations that redistribute the quantum projection noise. Previous analyses of potential metrological gain when using spin squeezing were performed on theoretically ideal states, without incorporating experimental imperfections or inherent limitations which result in non-unitary quantum state evolution. Here, we show that potential gains in clock stability are substantially reduced when the spin squeezing is non-unitary, and derive analytic formulas for the clock performance as a function of squeezing, excess spin noise, and interferometer contrast. Our results highlight the importance of creating and employing nearly pure entangled states for improving atomic clocks.
	\end{abstract}

\section{Introduction}

Spin squeezed states (SSSs) \cite{Kitagawa1993} offer a path toward entanglement-enhanced quantum sensors by reducing the variance of one spin quadrature. Typically, this potential improvement is quantified in terms of the metrological Ramsey squeezing parameter $\xi_R$ \cite{Wineland1994}, defined as $\xi^2_R = \frac{1}{C^2}\frac{\Delta S_{\min}^2}{S/2}$, where $\Delta S_{\min}^2$ is the smallest variance of any spin quadrature of the state, $S$ is the maximum possible length of the spin vector, and $C$ is the contrast of the complete Ramsey sequence. In this picture of quantum-enhanced Ramsey spectroscopy, spin squeezing reduces the measurement noise variance by a factor of $\frac{\Delta S_{\min}^2}{S/2}$ compared to the standard quantum limit (SQL) that can be attained in the absence of entanglement using a coherent spin state (CSS), while the $\frac{1}{C^2}$ term accounts for the reduction in squared signal due to interferometer contrast loss. However, the expression for $\xi_R$ does not account for other downsides in using SSSs in Ramsey spectroscopy, such as the increase in quantum noise (antisqueezing) in the conjugate spin direction \cite{Andre2004}.

Single-particle decoherence, i.e. uncorrelated noise between the particles, due to atom loss or spontaneous emission, typically has a more deleterious effect on SSSs than on uncorrelated collective atomic states. In particular, if the coherence properties of an atomic clock are limited by single-particle decoherence, squeezing is found to offer at best a small, and constant with atom number, improvement in ultimate clock stability \cite{Huelga1997,Ulam-Orgikh2001}. However, in state-of-the-art optical atomic clocks \cite{Bloom2014,Ushijima2015,Schioppo2017,Ludlow2015}, the dominant noise is not single-particle decoherence but rather phase noise in the local oscillator (LO) laser used to interrogate the narrow atomic transition. Even as LO laser technology improves \cite{Zhang2017}, there are many increasingly narrow atomic clock transitions \cite{Ludlow2006,Rosenband2007,Poli2008,Godun2014,Huntemann2014a} that would still leave LO stability as the primary limit to Ramsey time, and hence precision, in atomic clocks. 

\begin{figure}[tbhp]
	\centering
	\includegraphics[width = 1\columnwidth]{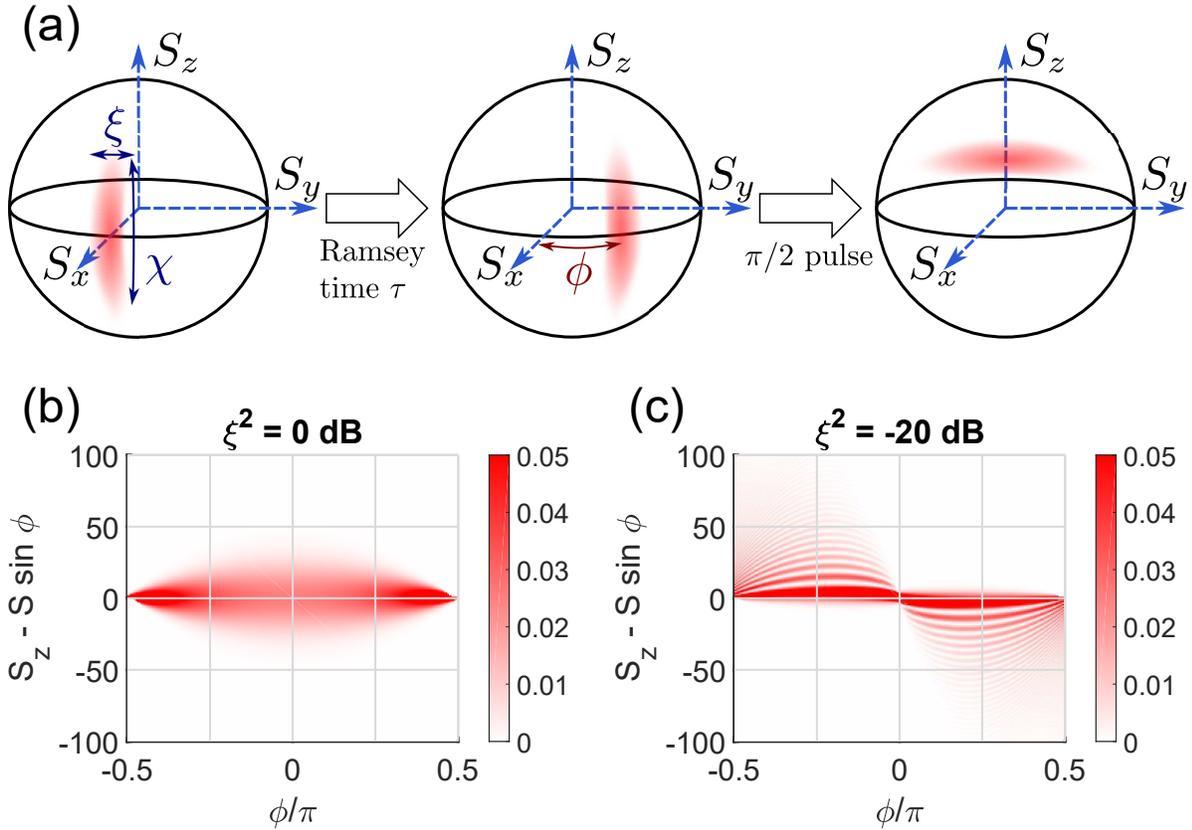}
	\caption{(a) Illustration of a Ramsey sequence using a spin squeezed state. A collective spin state pointing along $x$ is prepared with a quadrature reduced variance $\Delta S_y^2 = \frac{S}{2} \xi^2$ and an increased variance $\Delta S_z^2 = \frac{S}{2} \chi^2$. If the phase deviation $\phi$ is non-zero, the final readout of $S_z$ becomes sensitive to the antisqueezing $\chi$. (b) and (c) $S_z$ distributions, as a function of $\phi$ for a clock with $N=10^3$ spins, using a coherent state and a pure spin squeezed state with  $\xi^2 = \chi^{-2} = -20 \un{dB}$, respectively. Note that the $S_z$ distributions are vertically offset by the average clock signal $S \sin \phi$. The $S_z$ distribution is narrowed by squeezing only when $|\phi|$ is small.}
	\label{fig:RamseySpectroscopyWithSqueezedStates}
\end{figure}

In an atomic clock, the atomic phase is used to stabilize the LO phase. While dephasing of the LO does not destroy the quantum correlations between the atoms, it invalidates the assumption that the antisqueezing does not affect the measurement precision \cite{Andre2004}. The mechanism by which this happens is illustrated in Figure \ref{fig:RamseySpectroscopyWithSqueezedStates}(a). Due to the curvature of the Bloch sphere, part of the antisqueezed quadrature enters into the final $S_z$ measurement when the phase deviation $\phi$ between the atoms and the LO is non-zero, as in this case the final $\frac{\pi}{2}$ rotation of the Ramsey sequence places the SSS away from the equator of the Bloch sphere. Therefore, to avoid this leakage of antisqueezing into the final $S_z$ measurement, a squeezed clock must operate in a reduced range of $|\phi|$, limiting the Ramsey time $\tau$ in the presence of LO dephasing to a smaller value than what would be necessary to merely avoid $2\pi$ phase errors. On the other hand, the frequency stability of atomic clocks improves with longer Ramsey time as $\tau^{-1}$ for a single measurement, and as $\tau^{-1/2}$ for repeated measurements over a fixed total duration. Therefore, a conflict exists between obtaining the greatest phase noise suppression by using a short Ramsey time, and maximizing clock frequency stability with a long Ramsey time. Note that this effect does not harm sensors that operate in a fixed bandwidth, such as atomic magnetometers, accelerometers, and gyroscopes.

In recent years, many experiments have realized metrologically relevant entanglement using trapped ions \cite{Leibfried2004,Leibfried2005,Monz2011,Bohnet2016}, Bose-Einstein condensates (BECs) \cite{Gross2010,Hamley2012,Berrada2013,Strobel2014,Muessel2015}, and room-temperature \cite{Shah2010} and ultracold thermal atomic ensembles \cite{Appel2009,Leroux2010,Bohnet2014a,Cox2016a,Hosten2016}. Experimental realizations of spin squeezing are imperfect, and always produce states with more antisqueezing than squeezing ($\chi^2 > \xi^{-2}$), especially in low-density experiments suitable for clock operation \cite{Leroux2010,Bohnet2014a,Cox2016a,Hosten2016}. Although this excess antisqueezing has been noted experimentally, it has not been taken into account in previous studies of entanglement-enhanced clock stability \cite{Andre2004,Borregaard2013c,Borregaard2013b}.

Here, we extend the model of \cite{Andre2004} to derive analytical expressions including non-unitary squeezing, and determine the impact on potential gains in the stability of atomic clocks. We find that for typical states realized so far in experiments with dilute atomic ensembles amenable to clock applications \cite{Appel2009,Leroux2010,Bohnet2014a,Cox2016a,Hosten2016}, the impact can be severe, potentially wiping out any metrological gain from squeezing. We also analyze the effect of contrast loss, and find that the system is more robust to contrast loss during the squeezing than to contrast loss during the Ramsey time. We conclude that experimental efforts should be in part directed toward reducing excess antisqueezing, since moderate near-unitary squeezing can yield better clock stability than larger non-unitary squeezing.

%The simple picture claims that a state with $\xi^2$ less variance than the SQL will produce a measurement $\xi$ times more precise than with the equivalent measurement using any possible separable state, which holds if the presence of squeezing has no additional negative effects on the operation of the clock. However, we know this is not true -- entangled states are generally more fragile and subject to decoherence than separable states; and squeezing of one variance quadrature must inevitably introduce additional variance in the other noise quadrature, to satisfy the Heisenberg uncertainty principle.

% is most often framed as a way to obtain an improvement in the performance of quantum sensors, based on the potential reduction of quantum noise by entanglement. In this section, we would like to address the following question: in what situations is spin squeezing useful for enhancing the precision of atomic clocks?
 
%\section{Spin Squeezed Atomic Clock Model}
\section{Antisqueezing and Measurement Precision}

We consider an atomic clock based on a SSS consisting of $N = 2 S$ spin-$1/2$ atoms, with phase variance squeezed by a factor of $\xi^2\leq 1$ and population variance increased by a factor $\chi^2 \geq \xi^{-2}$. In terms of spin operator variances as shown in Figure \ref{fig:RamseySpectroscopyWithSqueezedStates}, this is equivalent to an initial phase squeezed state pointing along $S_x$, with $(\Delta S_y)^2 = \frac{S}{2} \xi^2$ and $(\Delta S_z)^2 = \frac{S}{2} \chi^2$. We quantify the imperfection of the squeezing process through the excess antisqueezing factor $A^2 = \xi^{2} \chi^{2}  \geq 1$, with equality corresponding to unitary squeezing (at contrast $C=1$). After the Ramsey time $\tau$, the atomic state is displaced in phase relative to the LO by the phase deviation angle $\phi$ due to LO noise. Contrast loss during state preparation or the Ramsey sequence arising from single-particle decoherence is neglected for the time being (but is discussed in Figure \ref{fig:ClockStabilityHosten2016} and Appendix \ref{sec:EffectOfContrastLoss}). The final $\pi/2$ rotation in Figure \ref{fig:RamseySpectroscopyWithSqueezedStates}(a) results in a distribution for $S_z$ that depends on the accumulated phase $\phi$, as shown in Figures \ref{fig:RamseySpectroscopyWithSqueezedStates}(b) and \ref{fig:RamseySpectroscopyWithSqueezedStates}(c). In a clock, the final measurement of $S_z$ is used to estimate $\phi$, and thereby apply feedback to stabilize the LO phase. For squeezed states, we note that for $\phi \neq 0$, the banana-like shape of the state wrapping around the Bloch sphere causes a leakage of the antisqueezed variance $\frac{S}{2} \chi^2$ into the final measurement of $S_z$, and therefore a deterioration in the ability to extract the true value of $\phi$ \cite{Andre2004}.
%The phase signal used to stabilize the LO to the atomic reference relies on the measurement of $S_z$ after the Ramsey time $\tau$ and the final $\pi/2$ rotation, from which we determine $\phi$. 
To determine the clock stability, we first calculate the variance $(\Delta \phi)^2$ of the phase measurement as a function of the squeezing $\xi$, antisqueezing $\chi$, and phase deviation $\phi$. We then use the derived analytical relation to determine the optimum Ramsey time $\tau$ and best attainable clock frequency stability.

The process by which $\phi$ is estimated is as follows. Given our knowledge of the initial quantum state, we calculate the conditional probability distribution $P(S_{z}|\phi)$ for each value of the true phase deviation $\phi$. With no prior knowledge about $\phi$, the Bayesian estimate of $\phi$, given a particular measured value $S_{z,f}$ of the $S_z$ operator at the conclusion of the Ramsey sequence, will be given by the conditional distribution $P(\phi | S_{z,f})$. In the limit where $P$ is Gaussian with respect to $S_{z}$ and $\phi$, the two distributions  $P(S_{z}|\phi)$ and  $P(\phi|S_{z})$ are directly related to one another through the signal slope $\frac{\partial \langle S_{z,f} \rangle}{\partial (\phi)}$, found from the mean signal produced by the clock sequence, $\langle S_{z,f} \rangle (\phi) = S \sin(\phi)$. Note that this last expression is only approximate for squeezed states, but remains sufficiently accurate for all $\phi$ of interest, as can be seen by comparing Figures \ref{fig:RamseySpectroscopyWithSqueezedStates}(b) and \ref{fig:RamseySpectroscopyWithSqueezedStates}(c).

%The process by which $\phi$ is estimated is illustrated in Figure \ref{fig:RamseySpectroscopyWithSqueezedStates}(b) and (c) for a CSS and a SSS with $\xi^2 = -15 \un{dB}$ respectively. For each value of the true phase deviation $\phi$, there is a corresponding distribution of final measurements $S_{z,f}$. Here, we can directly see what we qualitatively argued previously: the squeezed state offers a narrower distribution of $S_{z,f}$ only for sufficiently small $|\phi|$; outside of this range, the situation is in fact reversed.

%Conversely, when we measure a certain value for $S_{z,f}$, we can infer some knowledge about the phase, indicated by the distribution $P(\phi | S_{s,f})$. As long as the signal $S_{z,f}(\phi)$ is locally linear, these two distributions have a direct relationship through the signal slope. Therefore, to infer the precision of the phase measurement $\Delta_\phi^2(\tau)$, we need only to compute the width of the distribution for $S_{z,f}$.

%\begin{figure}[hbtp]
%	\centering
%	\includegraphics[width = 1.0\columnwidth]{QuantumClockIllustration.pdf}
%	\caption{Theoretical response curve for the measured atomic signal $S_z$ as a function of the phase deviation of the local oscillator $\phi$, for a clock using $1000$ atoms and (a) a coherent spin state (b) a squeezed spin state with $\xi^2 = -15\un{dB}$.}
%	\label{fig:RamseySpectroscopySignalAndNoise}
%\end{figure}

For the initial state in the Ramsey sequence, we already defined $(\Delta S_y)^2 = \frac{S}{2} \xi^2$ and $(\Delta S_z)^2 = \frac{S}{2} \chi^2$. The only remaining variance to compute is $(\Delta S_x)^2$. Using the Holstein-Primakoff approximation (see Appendix \ref{sec:DerivationOfDeltaSx}), we find
%In the Holstein-Primakoff approximation, we can consider a small region of the Bloch sphere as approximating a canonical $\{X,P\}$ phase space. Then, the $S_x$ quadrature corresponds to the excitation number operator $\hat{n}$. Using standard operator algebra, we can then find
\begin{equation}\label{eq:VarianceOfThirdQuadratureExact}
(\Delta S_x)^2 = \frac{(\chi^2-\chi^{-2})^2}{8}
\end{equation}
The variance of the final $S_z$ projection after the Ramsey sequence has two components: the initial $S_y$ variance, with weight $\cos^2(\phi)$, and the leaking in of $(\Delta S_x)^2$ into $S_z$, with weight $\sin^2(\phi)$. Putting these together gives
\begin{equation}\label{eq:FinalSzMeasurementNoise}
(\Delta S_{z,f})^2 = \frac{N}{4} \xi^2 \cos^2(\phi) + \frac{(\chi^2-\chi^{-2})^2}{8} \sin^2(\phi)
\end{equation}
The expected variance $(\Delta \phi)^2$ of the estimate for $\phi$ at the end of the Ramsey sequence is simply the variance of the final $S_z$ measurement given by (\ref{eq:FinalSzMeasurementNoise}), normalized by the slope of $\phi$ as a function of $S_{z,f}$: $(\Delta \phi)^2 = \left( \frac{\partial\langle S_z \rangle}{\partial\phi}\right)^{-2} \times (\Delta S_{z,f})^2$. Substituting (\ref{eq:FinalSzMeasurementNoise}) gives
\begin{equation}\label{eq:ClockStabilityAfterOneRamseySequence}
(\Delta\phi) ^2 = 
\frac{1}{N} \xi^2 + \frac{(\chi^2-\chi^{-2})^2}{2 N^2} \tan^2 (\phi)
\end{equation}
%This equation gives a great deal of insight into the performance of quantum sensors, both using classical and entangled states. For example, if we use a coherent spin state in our clock, we have 
%\begin{equation}
%\Delta_\phi^2(\tau) = 
%\left\langle
%\frac{1}{N C} + \frac{1-C}{N C^2} \sec^2 (\phi)
%\right\rangle_{\phi}
%\end{equation}
%If the Ramsey contrast is high ($C=1$), the performance of the sensor is independent of the phase shift itself: $\sigma_\phi^2(\tau) = \frac{1}{N}$. However, when contrast is lost, the sensor performance worsens not only because of a reduction in signal, but also because of an additional source of noise (the $\ket{\downarrow}$ atoms) that can significantly worsen the signal to noise ratio when $\phi$ is not near 0.

Note that (\ref{eq:ClockStabilityAfterOneRamseySequence}) is only valid when $|\phi|<\pi/2$, since that is the phase range in which the clock signal $\langle S_{z,f} \rangle (\phi)$ is invertible. Moreover, (\ref{eq:ClockStabilityAfterOneRamseySequence}) predicts a divergence of the phase error when $|\phi|=\pi/2$, which is an unphysical artifact of using a locally linearized model.

To estimate the maximum possible phase error $\phi _{max}$ near $|\phi|=\pi/2$, we note that if we measure a value of $S_z$ near the top (or bottom) of the Ramsey fringe, there is a finite range of $\phi$ which could have produced this value of $S_z$. The variance of $S_z$ at $|\phi|=\pi/2$ is given by 
$
(\Delta S_{z,f})^2 (\phi=\pi/2) = \frac{(\chi^2-\chi^{-2})^2}{8}
$. Therefore, the largest possible phase error $\Delta \phi_{max}$ must satisfy
$
\frac{N}{2} \left( 1-\cos(\Delta \phi_{max}) \right) = \Delta S_{z,f} (\phi=\pi/2)
$,
which gives, for $\Delta S_{z,f} \ll N/2$,
\begin{equation}
\label{eq:MaxPhaseErrorAtPi2}
\Delta \phi_{max} = \left[ 2 \frac{(\chi^2-\chi^{-2})^2}{N^2}  \right]^{1/4}
\end{equation}

To complete the picture, we need to estimate the phase error for $|\phi|>\pi/2$. In this case, the inversion function $\phi(S_{z,f})$ will give a result in the range $|\phi|<\pi/2$, with an error approximately equal to $4(|\phi|-\pi/2)^2$. To make the result continuous with the error in the vicinity of $|\phi|=\pi/2$, we approximate the phase error as
\begin{equation}
\label{eq:PhaseErrorOutsidePi2}
(\Delta \phi) ^2(|\phi|>\pi/2) = 4 (|\phi|-\pi/2)^2 + (\Delta\phi_{max})^2
\end{equation}
%This function grows linearly in error and quadratically in variance as we go away from $|\phi| = \pi/2$.

The complete dependence of the phase estimation error $(\Delta\phi)^2$ on $\phi$ can be obtained by stitching together (\ref{eq:ClockStabilityAfterOneRamseySequence}), (\ref{eq:MaxPhaseErrorAtPi2}), and (\ref{eq:PhaseErrorOutsidePi2}) for different values of $\phi$. To validate this approximate analytical formula, we perform a complete numerical simulation of the process of estimating a phase in a Ramsey sequence using states with different levels of squeezing, described in Appendix \ref{sec:NumericalSimulationsofDeltaPhi}. We find excellent agreement between these analytical approximations and the results of the full numerical simulation, especially around $\phi \approx 0$ where the phase estimation error is low, the crucial region for predicting optimal clock performance.

\section{Impact on Clock Stability}

Armed with an analytical model for the dependence of the phase estimation error $\Delta \phi$ on the phase difference $\phi$ between the LO and atomic ensemble given by (\ref{eq:ClockStabilityAfterOneRamseySequence}), we can now tackle the main question: How much can the frequency stability of a clock be improved by spin squeezing, in the limit where LO dephasing is the dominant noise source?

To quantify the precision of our clock, we need to evaluate the RMS difference between the true LO phase and the estimate of the LO phase we obtain using Ramsey spectroscopy of the atomic ensemble. We divide the total measurement time $T$ into $T/\tau$ intervals of duration $\tau$, corresponding to the individual Ramsey sequences of the experiment. 
%Then, the total phase error after a time $T$ will be given by
%\begin{equation}\label{eq:ClockStabilityAfterEntireExperiment}
%\sigma^2_{\phi}(T) = \frac{T}{\tau} \langle \phi^2 (\tau) \rangle
%\end{equation}
%But, as we can see from Figure \ref{fig:dPhiErrorNumericalVsAnalytical}, greater squeezing also reduces the region of  this is precisely what will happen when the LO phase has a longer time to diffuse around.
The overall variance of the LO phase estimate after a time $T$ is $T/\tau$ multiplied by the expected phase error variance after a single measurement,
\begin{equation}\label{eq:ClockStabilityWithGaussianNoise}
\sigma^2_{\phi}(T) = \frac{T}{\tau} \int_{-\infty}^{\infty} d\phi \, P(\phi,\tau) \left(\Delta \phi(\phi)\right)^2
\end{equation}
with corresponding frequency stability (or Allan deviation) given by $\sigma^2_{\omega} = T^{-2} \sigma^2_{\phi}$.
%Here, we see the trade-off that quantum-enhanced spectroscopy faces: a longer Ramsey time gives better clock performance, as long as $\langle \Delta\phi^2 (\tau) \rangle$ does not degrade too much. 

The probability distribution $P(\phi,\tau)$ of the LO having a phase deviation $\phi$ after Ramsey time $\tau$ depends on the linewidth and lineshape of the LO. % which play a key role in setting the clock stability.
As a simple example, we approximate the phase evolution of the LO as a Gaussian process, following the model of \cite{Andre2004}. For a free-running LO linewidth (or equivalently a dephasing rate) of $\gamma$, this model gives a Gaussian distribution for the phase deviation $\phi$, with variance 
$
\sigma_{LO} (\tau) = \gamma \tau
$
:
%where $\gamma$ is the dephasing rate of the LO, also equaling its free-running linewidth. 
\begin{equation}\label{eq:PhaseErrorAfterOneRamseyCycle}
P(\phi,\tau) = \frac{1}{\gamma \tau\sqrt{2 \pi}} \exp \left[ -\frac{\phi^2}{2 (\gamma \tau)^2} \right].
\end{equation}

%\section{Squeezed Clock Stability Limits}

\begin{figure}[hbtp]
	\centering
	\includegraphics[width = 1.0\columnwidth]{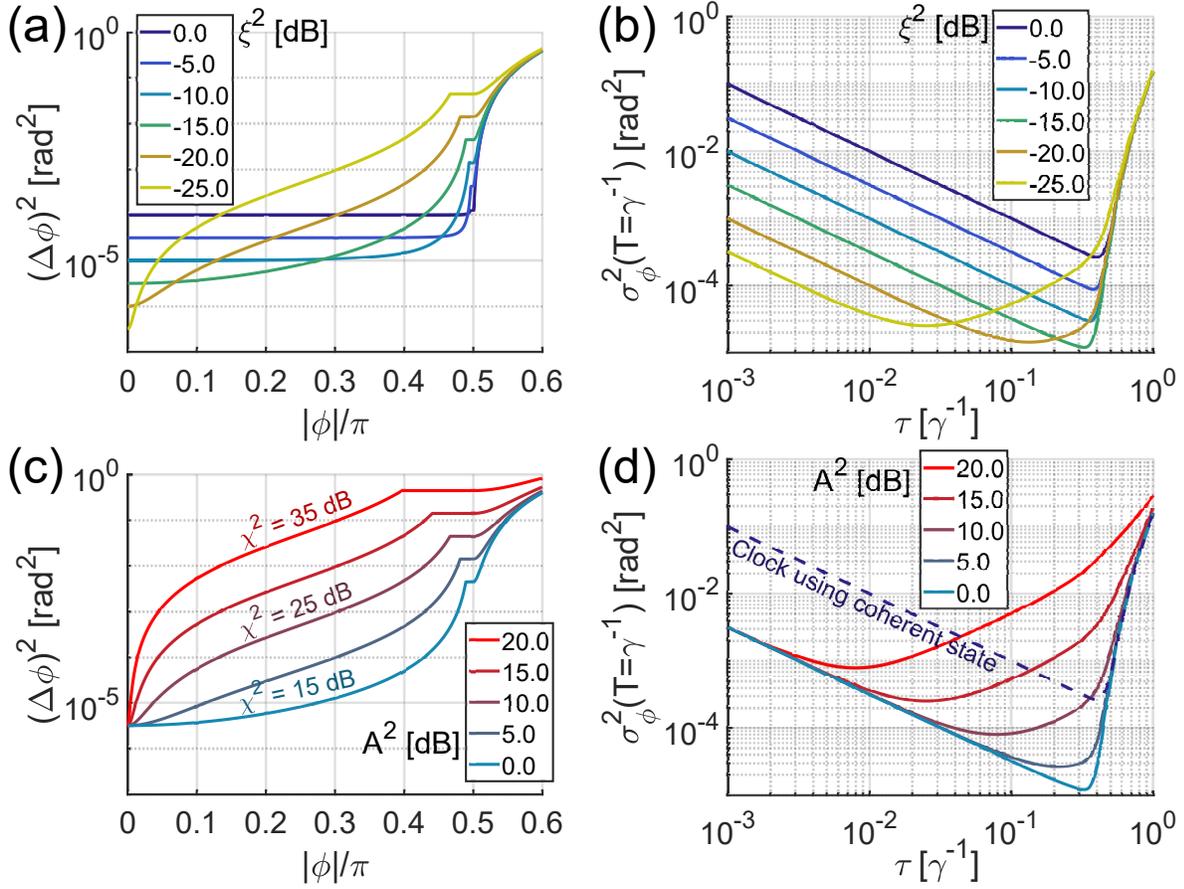}
	\caption{
		Phase estimation error and clock stability for squeezed states.
		(a) Phase estimation error $(\Delta \phi)^2$ after one Ramsey experiment for different amounts of unitary squeezing $\xi^2$ in a clock with $N=10^4$ atoms, as a function of the phase deviation $\phi$. (b) Corresponding clock stability after a time $T=\gamma^{-1}$, as a function of the Ramsey time $\tau$ (in units of $\gamma^{-1}$). (c) Same as (a), fixing the squeezing at $\xi^2 = -15 \un{dB}$, and varying the state area $A$ and accordingly the antisqueezing $\chi = A/\xi$. (d) Corresponding clock stability. Note that the steps in $(\Delta \phi)^2$ near $\frac{|\phi|}{\pi} \approx \frac{1}{2}$ are a consequence of transitioning between two regimes of approximating $(\Delta \phi)^2$ by (\ref{eq:ClockStabilityAfterOneRamseySequence}) and (\ref{eq:MaxPhaseErrorAtPi2}), and are not present in a full numerical simulation (shown in Figure \ref{fig:dPhiErrorNumericalVsAnalytical}). 
	}  
	\label{fig:PhaseErrorAndClockStability}
\end{figure}

Using this probability distribution, we calculate the phase estimation error $(\Delta\phi)^2$ and corresponding clock stability in several instructive cases, as shown in Figure \ref{fig:PhaseErrorAndClockStability}.

%The corresponding clock stability is shown in Figure \ref{fig:PhaseErrorAndClockStability}(b). As we can see, a longer Ramsey time leads to a better clock performance, up to a limit around $\gamma \tau\approx 0.45$, when the probability of large phase errors becomes significant. The best possible clock performance, optimized over $\tau$, scales as $\phi^2(T) \propto C^{-2}$, and reaches this optimum point at $\gamma \tau\approx 0.5$, regardless of the contrast $C$.

First, we consider a clock operating without excess antisqueezing. The corresponding phase estimation errors are shown in Figure \ref{fig:PhaseErrorAndClockStability}(a). We see that as the squeezing increases, the ability to resolve small rotations around $\phi=0$ improves: $\left(\Delta\phi(0)\right)^2=\xi^2/N$. However, this comes at a price: the range of phase deviations $\phi$ for which the measurement is better than the SQL is reduced. In terms of clock precision, as shown in Figure \ref{fig:PhaseErrorAndClockStability}(b), this effect manifests itself in a reduction of the optimal Ramsey time. 
%For example, for $-15 \un{dB}$ of squeezing, the best LO variance is about $10$ times better than for the case of no squeezing, and is attained for $\gamma \tau = 0.15$. The best possible clock performance becomes worse when the squeezing exceeds the optimal value of $\xi = N^{-1/4} = -20 \un{dB}$.
As a result, the best possible clock stability increases with squeezing until an optimum is reached near $\xi^2 = N^{-1/3}$, worsening again for larger values of squeezing, as found in \cite{Andre2004} for unitary squeezing.

% a somewhat longer optimal Ramsey time than found in \cite{Andre2004}

Next, we consider the effect of non-unitary squeezing, where the excess antisqueezing is a factor of $A^2 = \xi^2 \chi^2$ in variance. We fix $\xi^2 = -15 \un{dB}$ and vary the antisqueezing $\chi = A/\xi$. The error in the phase measurement is shown in Figure \ref{fig:PhaseErrorAndClockStability}(c). As expected, excess antisqueezing has no effect on the measurement precision near $\phi=0$, but makes the error $\Delta \phi$ significantly worse for larger $|\phi|$. Turning to the clock stability in Figure \ref{fig:PhaseErrorAndClockStability}(d), we see that squeezed states with excess antisqueezing lead to clocks with the same stability for short Ramsey times, but the stability saturates at much smaller values of $\gamma\tau$.
As a result, when the state area exceeds $A^2 = 15\un{dB}$ (for $N=10^{4}$ atoms), the best attainable long-term stability of the squeezed clock is worse than if we had simply used a coherent spin state.
% For example, a state with an area of $A=4$ (corresponding to squeezing of $\xi^2 = -10 \un{dB}$ and antisqueezing $\chi^2 = 22.04 \un{dB}$), the best clock stability is obtained for $\gamma \tau = 0.09$, and is almost exactly the same as a clock with no squeezing. Thus, we can see that even a modest level of excess antisqueezing can completely negate any clock performance benefit that may be gained from squeezing.

\begin{figure}[tbp]
	\centering
	\includegraphics[width = 0.8\columnwidth]{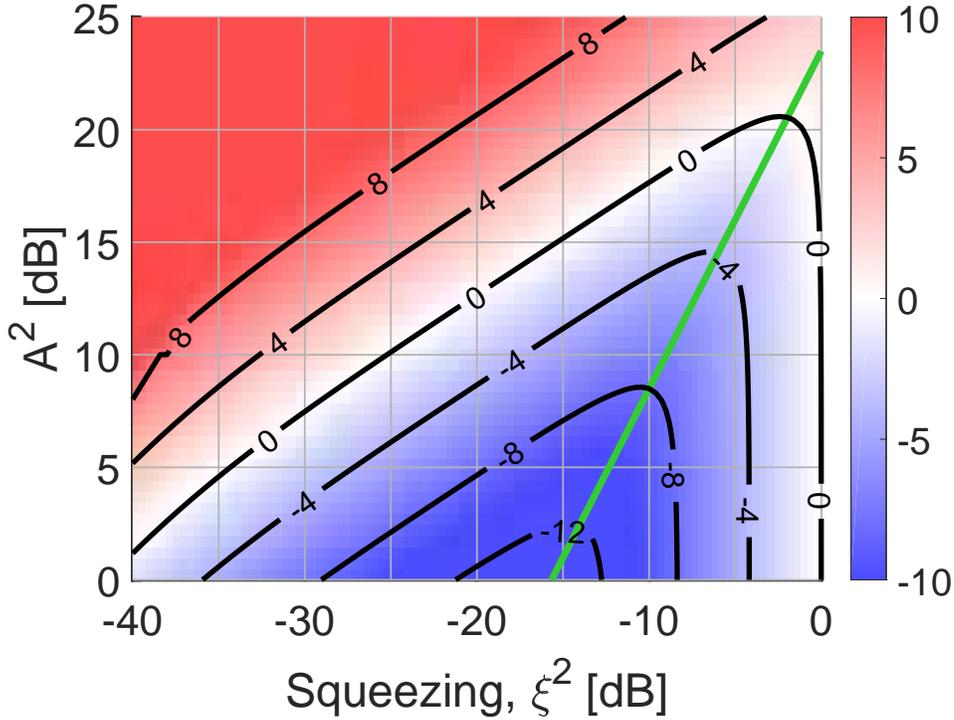}
	\caption{
		Clock performance in the presence of excess antisqueezing.
		Best possible clock stability, in units of $\mathrm{dB}$ of clock phase variance compared to a clock at the SQL, as a function of the squeezing $\xi^2$ and excess antisqueezing $A^2$, for $N=10^4$ atoms. The green line at $A^4 \xi^{-6} = 5 N$ indicates the transition between the regimes where clock performance improves or deteriorates with increasing squeezing.
	}  
	\label{fig:BestClockStabilityVsSqueezingAndAntisqueezing}
\end{figure}

Figure \ref{fig:BestClockStabilityVsSqueezingAndAntisqueezing} shows clock stability for an optimized Ramsey time $\tau$, as a function of squeezing $\xi^2$ and excess antisqueezing $A^2$. There are two distinct regions in the plot, delineated by which term in (\ref{eq:ClockStabilityAfterOneRamseySequence}) is greater. When $A^4 \xi^{-6} N^{-1} \ll 1$, i.e. when $\xi^2 > N^{-1/3}$ and the antisqueezing is moderate, the antisqueezing term in (\ref{eq:ClockStabilityAfterOneRamseySequence}) is small for all $|\phi| < \pi/2$. In this regime, the antisqueezing plays no role in determining the optimal Ramsey time or clock stability, with the latter improving in direct proportion to $\xi$, independently of $A$: $\sigma^2_{\phi}(\gamma^{-1}) = \xi^2 N^{-1}$. In the other limit, $A^4 \xi^{-6} N^{-1} \gg 1$, the optimal Ramsey time becomes shorter, with $\tau_{opt} = \gamma^{-1} \sqrt{N} \xi^3 A^{-2}$, yielding a clock stability of $\sigma^2_{\phi}(\gamma^{-1}) = A^2 \xi^{-1} N^{-3/2}$. Note that in this regime, the clock performance deteriorates with increasing squeezing. Numerically, we find the boundary of the two regions to lie near $A^4 \xi^{-6} N^{-1} = 5 $, in agreement with the result $\xi^2 \propto N^{-1/3}$ for optimum unitary squeezing ($A=1$) found in \cite{Andre2004}.

%Finally, we consider the effect of non-unitary squeezing on the performance of a squeezed clock. Figure \ref{fig:BestClockStabilityVsSqueezingAndAntisqueezing} shows the effect of excess antisqueezing on clock performance, again normalized to a clock with no squeezing and perfect contrast. We see that the potential clock gain is strongly reduced when the squeezing is non-unitary: for example, a state area of $A=2$, corresponding to $6 \un{dB}$ more antisqueezing than squeezing, limits the potential improvement in clock performance to $-5.8 \un{dB}$, attained at a squeezing of $\xi^2 = -8.6 \un{dB}$. If the state area exceeds $N^{-1/4} = 5.6$, no metrological gain can be obtained compared to a clock operating with a CSS, regardless of the amount of squeezing.

\begin{figure}[tbp]
	\centering
	\includegraphics[width = 1\columnwidth]{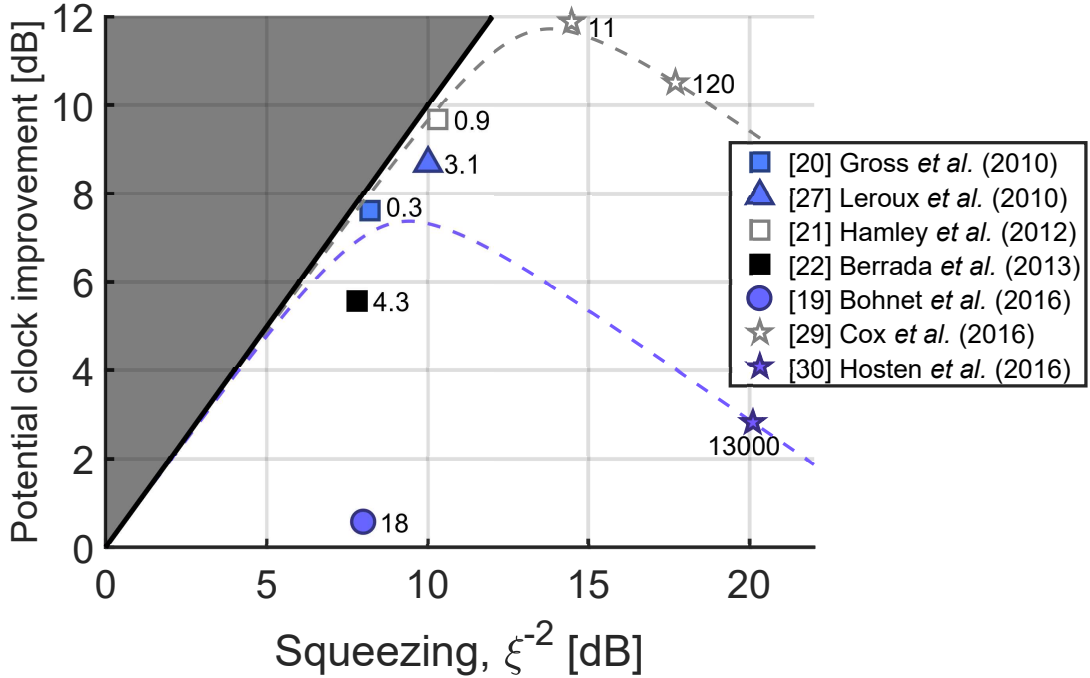}
	\caption{
		Potentially realizable clock stability gain using experimentally generated squeezed states after subtraction of detection noise, compared to a clock at the SQL. Each point reflects the reported squeezing assuming perfect state detection and Ramsey contrast $C=1$. Circles, squares, and triangles correspond to squeezing via collective interactions in trapped ions, Bose-Einstein condensates, and ultracold atomic ensembles respectively, while stars correspond to measurement-based squeezing with ultracold atomic ensembles. 
		Dashed lines extrapolate the results of \cite{Cox2016a,Hosten2016}, assuming a squeezing-independent state area $A$, as is typical for measurement-based squeezing experiments.
		Numerical labels indicate the value of $\alpha = A^4 \xi^{-6} N^{-1}$ for each experimental result. When $\alpha > 1$, the state has too much squeezing or excess antisqueezing, preventing the full utilization of available squeezing in boosting clock performance.
	}  
	\label{fig:ClockStabilityRecentSqueezing}
\end{figure}

Using this approach, we can analyze the potential gains in clock stability that could be obtained using spin squeezed states that have been experimentally realized. As shown in Figure \ref{fig:ClockStabilityRecentSqueezing}, better squeezing $\xi^2$ does not necessarily lead to greater clock stability, especially if it is achieved at the expense of excess antisqueezing. Experiments where the parameter $\alpha =  \frac{A^4}{\xi^{6}N}$ is less than one, such as \cite{Berrada2013,Gross2010,Leroux2010,Hamley2012}, would be able to employ the full amount of generated squeezing in improving an atomic clock. On the other hand, experiments with $\alpha \gg 1$, including those with the largest observed squeezing \cite{Bohnet2016,Hosten2016,Cox2016a}, produce a situation where this squeezing cannot be used in such an efficient manner. It is interesting to note that \cite{Berrada2013,Gross2010,Leroux2010,Hamley2012} all employ collective interactions to generate squeezing, while \cite{Hosten2016,Cox2016a} use QND measurements for entangling the atoms. In measurement-based squeezing, any undetected photons produce excess antisqueezing; to minimize the latter, one needs to maximize the quantum efficiency of light detection and optimally use the available information in the probe light. This suggests that even though measurement-based squeezing has been used to create smaller spin variances, collective atom-atom interactions \cite{Thomsen2002,Schleier-Smith2010,Lewis-Swan2018} may offer better performance for spin squeezed clocks.

\begin{figure}[tbp]
	\centering
	\includegraphics[width = 1.0\columnwidth]{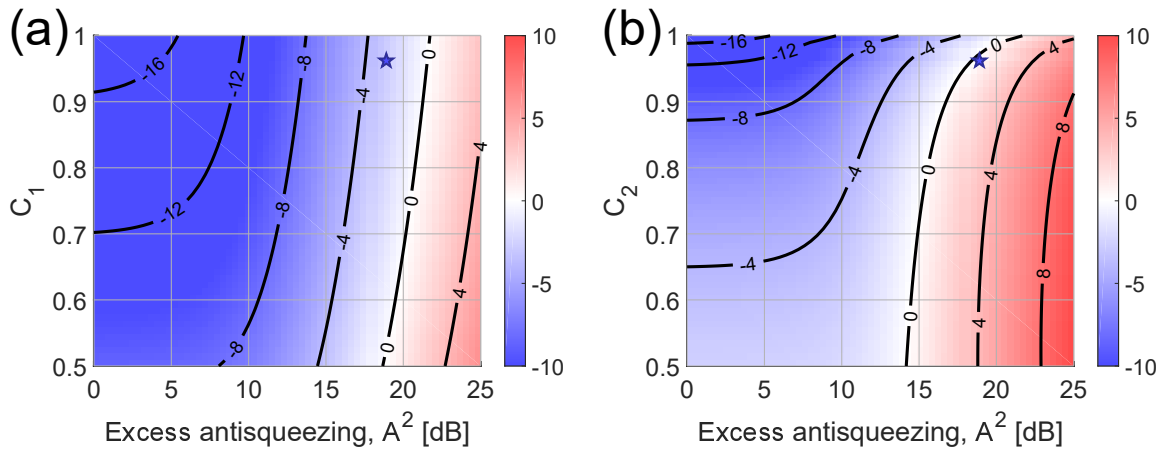}
	\caption{
		Effect of contrast loss on clock performance.
		Possible stability improvement (in units of dB) of a spin squeezed clock over a coherent spin state clock, as a function of excess antisqueezing $A^2$ and (a) squeezed state generation contrast $C_1$ and (b) Ramsey time contrast $C_2$. We consider a squeezed state with $N=5\times10^5$ atoms and squeezing $\xi^2 = -20.1 \un{dB}$ as realized in \cite{Hosten2016}. The star indicates the experimentally observed values of $A^2$ and $C$. In both plots, we assume unity state preparation contrast $C_1 = 1$ for the CSS. In (b), we vary $C_2$ for both the CSS and SSS clocks, since contrast loss during the Ramsey time is unaffected by the squeezing process and should be equal for coherent and spin squeezed states.
	}  
	\label{fig:ClockStabilityHosten2016}
\end{figure}

In practice, Ramsey contrast in clocks is usually below unity, due to atom loss or technical imperfections. By combining the effects of antisqueezing and Ramsey contrast decay, we can readily determine parameter regions of SSSs that can enhance atomic clocks, as discussed in Appendix \ref{sec:EffectOfContrastLoss}. The total Ramsey contrast $C = C_1 \times C_2$ has two components. The first, $C_1$, is the reduction in contrast acquired during the preparation of the SSS, which will be present in optically induced spin squeezing due to the inevitable scattering of probe photons by the atoms during state preparation. The second factor, $C_2$, is contrast loss during the Ramsey time itself. 
% limits clock performance over the SQL to a factor of $\frac{1}{1-C}$, due to uncorrelated spin noise from decohered atoms. 

An example of the effect of contrast loss is shown in Figure \ref{fig:ClockStabilityHosten2016}, where we analyze the experiment of \cite{Hosten2016}, which reported the creation of a state with $\xi^2 = -20.1 \un{dB}$, $A^2 = 19 \un{dB}$, and $C=0.962$ for $N=5\times10^5$ atoms. For these parameters, a change in clock stability ranging between an improvement by $2.7 \un{dB}$ and a deterioration by $0.6\un{dB}$ could be expected, as compared to a clock operating with a CSS, depending on where in the Ramsey sequence the contrast loss originates. 
We see that contrast loss during spin squeezing, shown in Figure \ref{fig:ClockStabilityHosten2016}(a), is relatively benign in terms of clock stability compared to contrast loss during the Ramsey time, as shown in Figure \ref{fig:ClockStabilityHosten2016}(b); this is also noted in \cite{Appel2009} and further discussed in Appendix \ref{sec:EffectOfContrastLoss}.
For the SSS of \cite{Hosten2016}, if the excess antisqueezing could be reduced to $A^2<7\un{dB}$ while maintaining $C_1>0.7$, $C_2>0.96$, then the clock could be operated $10\un{dB}$ below the SQL.
%In that experiment, long-term clock stability can be improved by $10 \un{dB}$ only by staying in a narrow parameter regime of $C>0.95$ and $A^2 < 7 \un{dB}$. 
Thus, practical atomic clocks must control both excess antisqueezing and contrast loss at a tight level to profit from spin squeezing.

\section{Conclusion}

We have analyzed the effect of non-unitary squeezing and derived simple expressions for potential improvements in clock stability in the presence of LO noise, as a function of the squeezing and anti-squeezing. We find that for a state with $N$ atoms, the squeezed state offers no metrological gain over a coherent spin state if the excess antisqueezing variance exceeds $N^{-1/2}$, and that highly squeezed states with large excess antisqueezing can lead to worse clock performance than moderately squeezed states with less antisqueezing. Therefore, experiments aiming for ultimate clock stability will benefit from operating close to the fundamental limit of unitary squeezing, by squeezing with collective atom-atom interactions, or by reducing light loss in measurement based squeezing.

We emphasize that these results apply to atomic clocks but not to sensors that require signal readout after an externally-imposed Ramsey interrogation time, such as those used to measure a time-dependent signal. In this case, for short interrogation times, it is the spin squeezing alone that determines the sensor performance.

We would also like to note that schemes to extract the full metrological gain from squeezed states in the presence of LO noise have been proposed, by using ensembles of clocks \cite{Borregaard2013c}, or by measurement and active feedback onto the atomic state \cite{Borregaard2013b}. It remains to be analyzed how much these approaches can enhance the stability of clocks in the presence of non-unitary spin squeezing and other experimental imperfections.

\ack

This work was supported by NSF grants \#PHY-1505862 and \#PHY-1806765, the Center for Ultracold Atoms NSF grant \#PHY-1734011, and ONR grant \#N00014-17-1-2254.

\appendix

\section{\label{sec:DerivationOfDeltaSx} Derivation of Expression for $\Delta S_x^2$}

In the Holstein-Primakoff approximation \cite{Holstein1940} of an N-particle Bloch sphere as a tangent plane perpendicular to the $+\hat{x}$ direction, we have the following mapping between harmonic oscillator operators $a$, $a^\dagger$ and spin operators:
\begin{eqnarray}
S_x &= \frac{N}{2} - a^\dagger a \nonumber\\
S_y &=  \frac{\sqrt{N}}{2 i} (a^\dagger - a)\\
S_z &= \frac{\sqrt{N}}{2} (a^\dagger + a) \nonumber
\end{eqnarray}
The harmonic oscillator ground state $\ket{0}$ corresponds to an atomic coherent state pointing in the $+\hat{x}$ direction:
\begin{equation}
\ket{+\hat{x}}^{\otimes N} = \ket{0} \nonumber
\end{equation}
The squeezing and displacement operators are defined in the usual way:
\begin{eqnarray}
S(\lambda) &= \exp \left( \frac{1}{2}\left( \lambda^* a^2 - \lambda {a^\dagger}^2 \right)  \right)  \nonumber\\
D(\alpha) &= \exp \left( \alpha a^\dagger - \alpha^* a  \right)\nonumber
\end{eqnarray}
When $\lambda$ is real, the state $S(\lambda) \ket{0}$ is squeezed in the $S_y$ quadrature with variance $\xi^2 = \frac{S}{2} e^{-2 \lambda}$ and antisqueezed in the $S_z$ quadrature with variance $\chi^2 = \frac{S}{2} e^{2\lambda}$. 

A non-unitary squeezed state with antisqueezing variance $\chi^2$ and squeezing variance $\xi^2$ can be decomposed as a statistical mixture of unitary squeezed states with squeezing $\xi_0^2 = \chi^{-2}$, displaced in the $S_y$ direction with a Gaussian probability density of variance $\xi^2 - \xi_0^2$. For a squeezed state $\ket{\psi} =  D(\alpha) S(\lambda) \ket{0}$, we find the following expectation values for the number operator $n = a^\dagger a$:
\begin{eqnarray}
\braket{n} &= \sinh^2 (\lambda) + |\alpha|^2 \nonumber\\
\braket{n^2} &= \braket{n}^2 + 2 \sinh^2 (\lambda) \cosh^2 (\lambda) + \\ & + |\alpha|^2 \left(\sinh^2 (\lambda) + \cosh^2 (\lambda) \right) + \nonumber \\ &+ \left(\alpha^2 + {\alpha^*}^2\right) \sinh(\lambda) \cosh(\lambda) \nonumber
\end{eqnarray}

A displacement operator in the $S_y$ direction corresponds to a purely imaginary value for $\alpha$. Therefore, $(\Delta S_x)^2 = \braket{n^2} - \braket{n}^2 $ equals
\begin{equation}
\left(\Delta S_x  (\alpha,\lambda)\right)^2 = \frac{(e^{2\lambda} - e^{-2\lambda})^2}{8} + |\alpha|^2 \nonumber
\end{equation}
Evaluating the expectation value over the $\alpha$ distribution and substituting $\chi$ and $\xi$ gives the final result:
\begin{equation} \label{eq:DeltaSxSquaredDerivation}
\left(\Delta S_x (\xi,\chi)\right)^2  = \frac{(\chi^2 - \chi^{-2})^2}{8} + \xi^2 - \chi^{-2}
\end{equation}
For any squeezed state, we have $0\leq\xi^2 - \chi^{-2} < 1$, which will have negligible impact on the phase estimate precision $\Delta \phi$. Therefore, we can simplify (\ref{eq:DeltaSxSquaredDerivation}) to give
\begin{equation} \label{eq:DeltaSxSquaredDerivationSimple}
(\Delta S_x)^2 = \frac{(\chi^2 - \chi^{-2})^2}{8}
\end{equation}

\section{\label{sec:NumericalSimulationsofDeltaPhi} Numerical Validation of Phase Variance Formulas}

\begin{figure}[hbtp]
	\centering
	\includegraphics[width = 0.8\columnwidth]{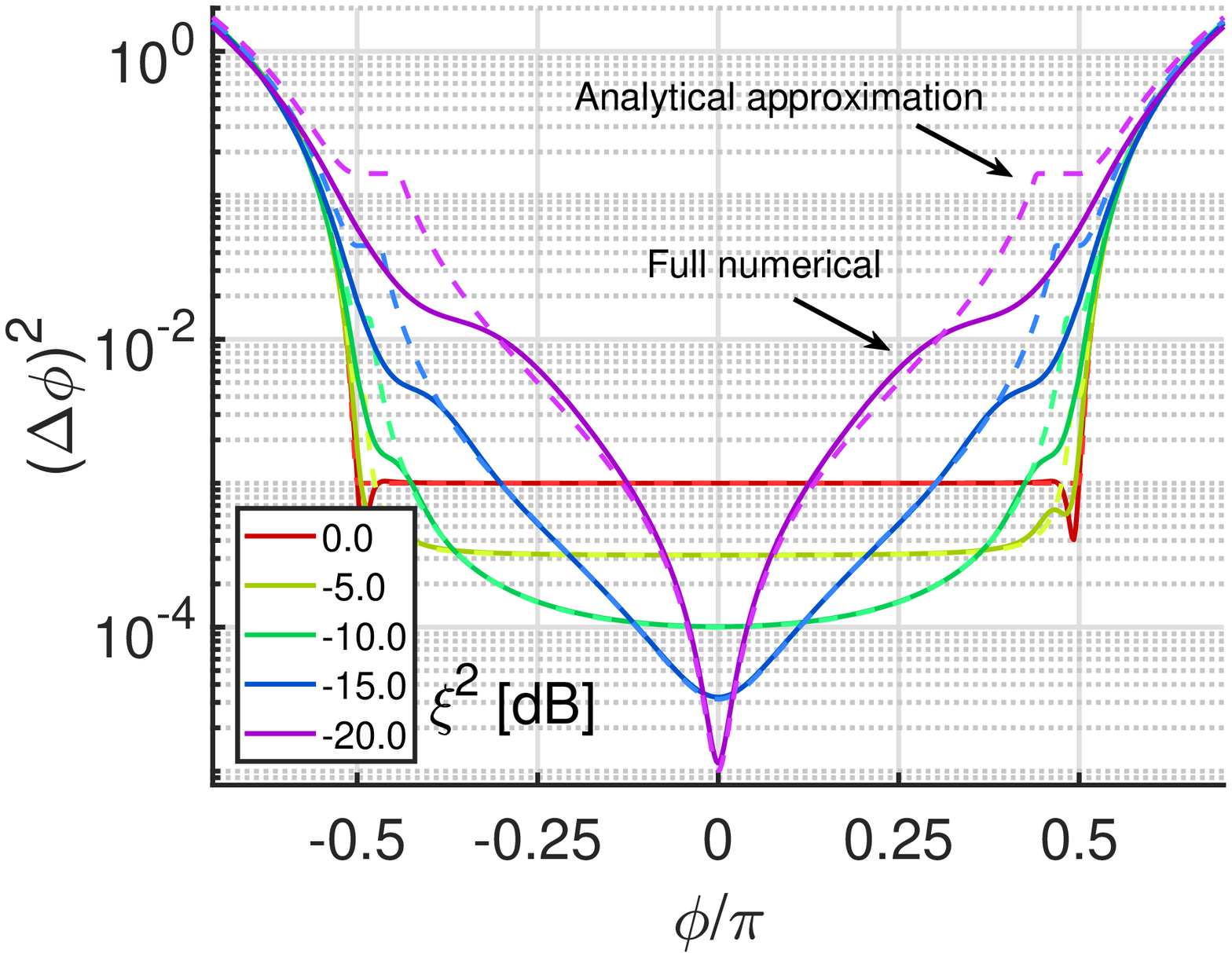}
	\caption{Expected error $(\Delta \phi)^2$ in the estimate of the LO phase $\phi$ as a function of $\phi$ for different amounts of squeezing $\xi$ and no excess antisqueezing, for $N=10^3$ spins. Dashed lines are the analytical formula built from (\ref{eq:ClockStabilityAfterOneRamseySequence}), (\ref{eq:MaxPhaseErrorAtPi2}), and (\ref{eq:PhaseErrorOutsidePi2}), while the solid lines are the result of a numerical simulation.}  
	\label{fig:dPhiErrorNumericalVsAnalytical}
\end{figure}

Here, we describe the procedure for validating the analytical formulas (\ref{eq:ClockStabilityAfterOneRamseySequence}), (\ref{eq:MaxPhaseErrorAtPi2}), and (\ref{eq:PhaseErrorOutsidePi2}) through a numerical simulation and generating the results in Figure \ref{fig:dPhiErrorNumericalVsAnalytical}. For each value of unitary squeezing $\xi$ and LO-atom phase deviation $\phi$, we calculate the probability distribution $P(S_z,\phi,\xi)$ of obtaining a particular value of $S_z$ during the projective measurement at the end of the Ramsey sequence. From this probability distribution, we determine the estimate for $\phi$ with the smallest root-mean-square error, given a particular observed value for $S_z$, assuming a uniform Bayesian prior for $\phi$ in the interval $(-\pi/2,\pi/2)$. This estimate is given by
\begin{equation*}
\phi_{est}(S_z,\xi) = \frac{\int_{-\pi/2}^{\pi/2} d\phi \, P(S_z,\phi,\xi) \times \phi}{\int_{-\pi/2}^{\pi/2} d\phi \, P(S_z,\phi,\xi)}
\end{equation*}
Next, for each true value of $\phi$ (which can lie outside the $(-\pi/2,\pi/2)$ range), we evaluate the expected error in the $\phi$ estimate:
\begin{equation}
\left(\Delta \phi(\phi)\right)^2 = \frac{\sum_{S_z = -N/2}^{N/2} P(S_z,\phi,\xi) \times (\phi_{est}(S_z,\xi)-\phi)^2}{\sum_{S_z = -N/2}^{N/2} P(S_z,\phi,\xi)}
\end{equation}
This is the function plotted in Figure \ref{fig:dPhiErrorNumericalVsAnalytical}. The effects of contrast loss and excess antisqueezing can be introduced by the appropriate modification of the function $P(S_z,\phi,\xi)$ to yield similarly good agreement with the analytical estimates of $(\Delta\phi)^2$ derived above.

\section{\label{sec:EffectOfContrastLoss} Effect of Contrast Loss}

\begin{figure}[tbhp]
	\centering
	\includegraphics[width = 1\columnwidth]{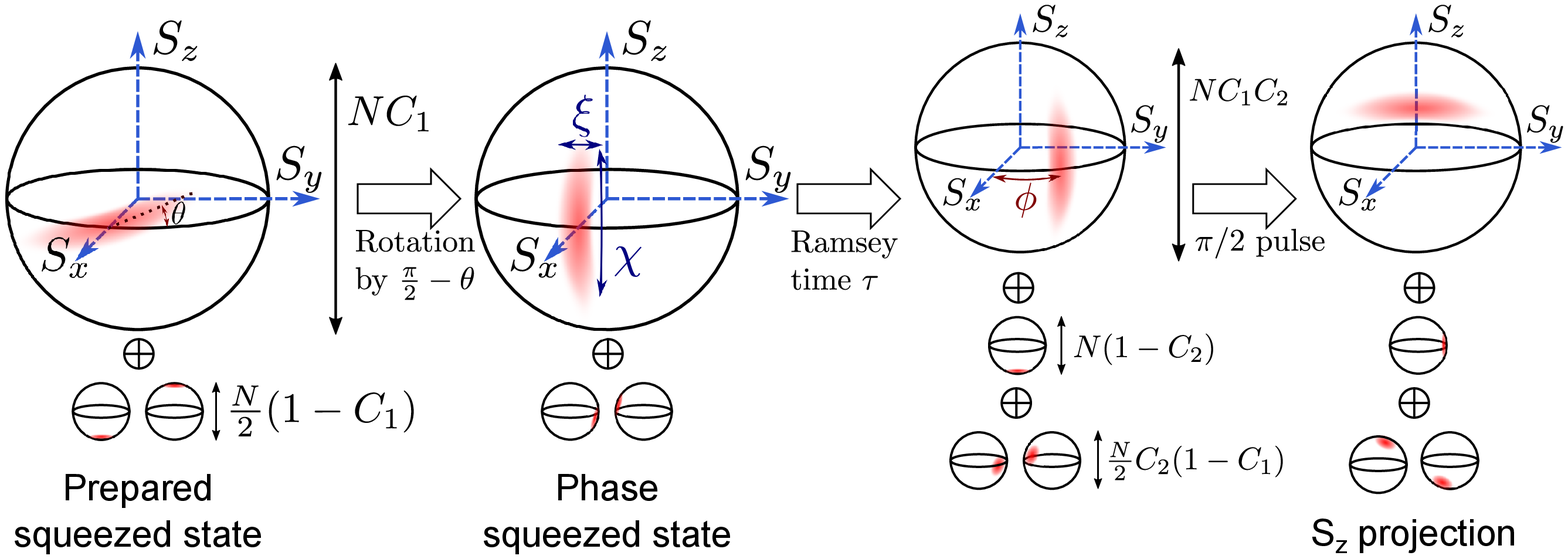}
	\caption{
		Ramsey sequence including the effect of contrast loss due to imperfection in spin squeezing generation ($C_1$) and single-particle decay into the ground state ($C_2$).
	}  
	\label{fig:RamseySpectroscopyWithSqueezingAndContrastLoss}
\end{figure}

The model we use can also be extended to include the effects of contrast loss, as shown in Figure \ref{fig:RamseySpectroscopyWithSqueezingAndContrastLoss}. We model contrast loss during the SSS preparation by combining a SSS comprised of $NC_1$ atoms with two equally populated sub-ensembles, one with  $\frac{N}{2} (1-C_1)$ atoms in the ground state, and the other with the same number of atoms in the excited state, giving a total Ramsey contrast of $C_1$.  
%With $N$ total atoms, a SSS comprised of $N C_1$ atoms and two CSSs with population $\frac{N}{2} (1-C_1)$ each give a total Ramsey contrast of $C_1$. 
The orientation of the SSS relative to the $S_z = 0$ plane depends on the method by which squeezing is produced -- measurement-based squeezing \cite{Bohnet2014a,Cox2016a,Hosten2016} gives $\theta = 0$ while feedback-based squeezing \cite{Leroux2010,Schleier-Smith2010,Thomsen2002} has $\theta = \arcsin (1/\chi)$.

During the Ramsey sequence, additional contrast loss by a factor $C_2$ may occur when some of the atoms decay from the excited clock state. We model this effect by a population transfer to a CSS containing $N (1-C_2)$ atoms in the ground state. The net Ramsey contrast of the entire sequence equals $C = C_1 \times C_2$, as only the atoms remaining in the SSS are still contributing to the signal.

Contrast loss adversely affects the clock performance by reducing the magnitude of the signal by a factor $C$, and by adding additional noise in the final $S_z$ measurement:
\begin{eqnarray}
(\Delta S_{z,CSS})^2 &= \frac{N C_2 (1-C_1)}{4} \left(1 - \cos^2 (\phi) \cos^2 (\theta) \right)\nonumber\\
 &+ \frac{N (1-C_2)}{4}
\end{eqnarray}
with the first term arising from contrast loss during squeezing preparation, and the second term from contrast loss during the Ramsey sequence. In typical situations where $\theta$ and $\phi$ are small, the second term is greater, reflecting the fact that contrast loss during the Ramsey sequence is a more severe problem than contrast loss during the squeezing preparation. These noise terms are added to the noise arising from the spin squeezed state itself, giving
\begin{eqnarray}\label{eq:FinalSzMeasurementNoiseWithContrastLoss}
(\Delta S_{z,f})^2 &= C \times \left[ \frac{N}{4} \xi^2 \cos^2(\phi) + \frac{(\chi^2-\chi^{-2})^2}{8} \sin^2(\phi) \right] + \nonumber \\ &+ \frac{N}{4} \times \left[ 1 - C - C_2 (1-C_1) \cos^2 (\phi) \cos^2 (\theta) \right]
\end{eqnarray}

The mean signal produced by the clock sequence is $\langle S_{z,f} \rangle (\phi) = C \times S \sin(\phi)$, giving the full expression for $(\Delta\phi)^2$:
\begin{eqnarray}\label{eq:ClockStabilityAfterOneRamseySequenceWithContrastLoss}
\left(\Delta \phi\right)^2 &= 
\frac{1}{C N} \xi^2 + \frac{(\chi^2-\chi^{-2})^2}{2 C N^2} \tan^2 (\phi) \nonumber \\
 &+ \frac{1-C}{C^2 N} \sec^2(\phi) - \frac{C_2(1-C_1)}{C^2 N} \cos^2(\theta)
\end{eqnarray}
Similarly, the maximum phase estimation error becomes
\begin{equation}
\label{eq:MaxPhaseErrorAtPi2WithContrastLoss}
\Delta \phi_{max} = \left[ 2 \frac{(\chi^2-\chi^{-2})^2}{N^2 C} + 4 \frac{1-C}{NC^2} \right]^{1/4}
\end{equation}

By replacing (\ref{eq:ClockStabilityAfterOneRamseySequence}) and (\ref{eq:MaxPhaseErrorAtPi2}) with (\ref{eq:ClockStabilityAfterOneRamseySequenceWithContrastLoss}) and (\ref{eq:MaxPhaseErrorAtPi2WithContrastLoss}), we can evaluate the performance of clocks in the presence of contrast loss and potentially non-unitary squeezing. 
%Note that contrast loss through single-particle decoherence that occurs while the state is number squeezed has a smaller impact on clock performance than if this loss occurs while the state phase squeezed, the situation analyzed here.

\begin{figure}[tbp]
	\centering
	\includegraphics[width = 1.0\columnwidth]{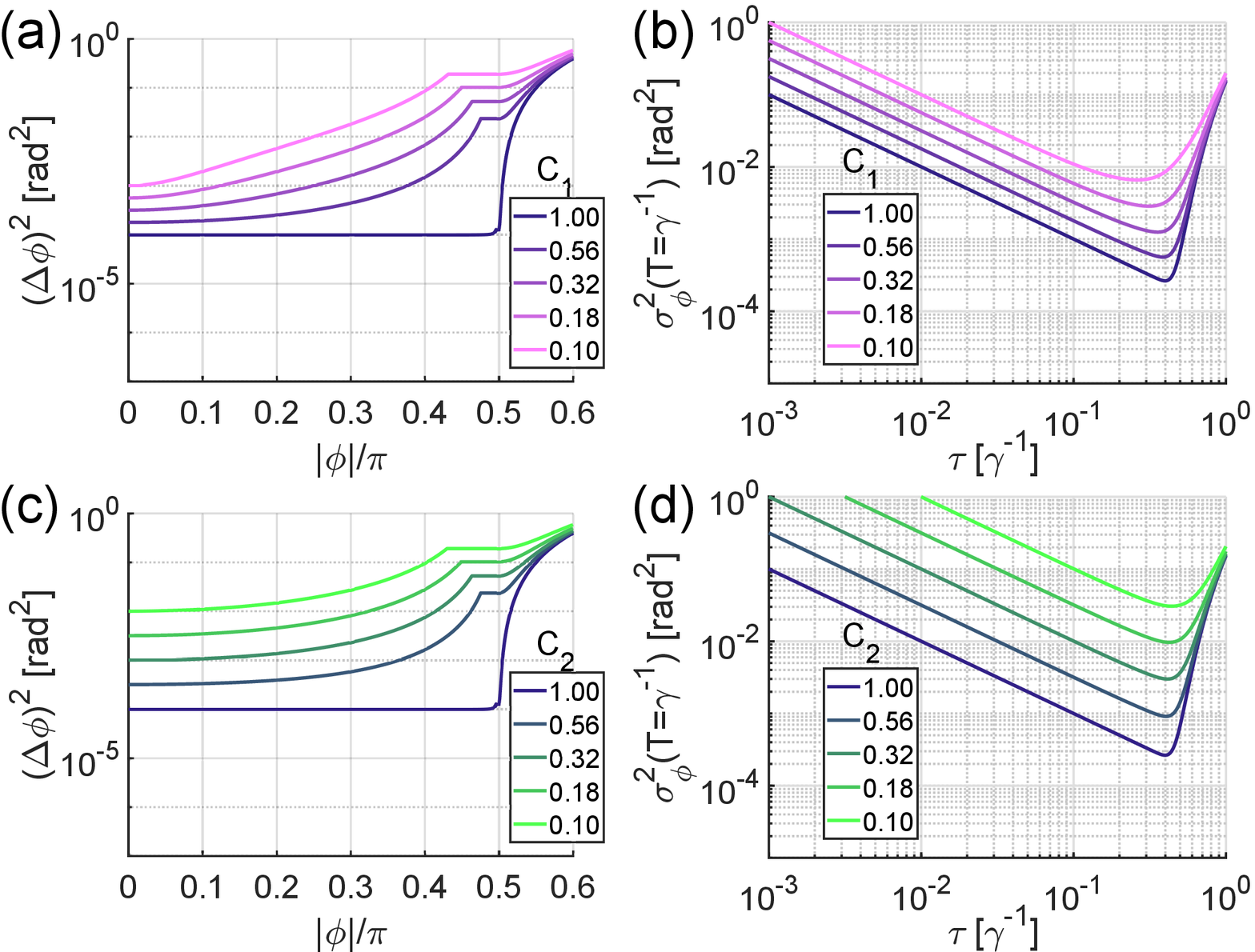}
	\caption{
		(a) Phase error $(\Delta \phi)^2$ in a clock with $N=10^4$ for different preparation contrast $C_1$, no squeezing, and no contrast loss during the Ramsey sequence, as a function of the phase deviation $\phi$. (b) Corresponding clock stability after a time $T=\gamma^{-1}$. (c) Same as (a), but with perfect state preparation contrast $C_1=1$, and varying amounts of contrast $C_2$. (d) Corresponding clock stability.
	}  
	\label{fig:PhaseErrorAndClockStabilityWithContrastLoss}
\end{figure}

As an example, consider a clock operating without any squeezing, for different amounts of remaining contrast $C$ after the Ramsey time $\tau$. This contrast loss may occur during the state preparation ($C_1 < 1$, $C_2 = 1$) or during the Ramsey sequence itself ($C_1 = 1$, $C_2<1$). The errors in estimating the LO phase for these two situations are shown in Figure \ref{fig:PhaseErrorAndClockStabilityWithContrastLoss}(a) and (c). When $C=1$, the error in phase estimation is always at the SQL for $N=10^4$ atoms: $(\Delta \phi)^2=10^{-4}=1/N$. As the contrast becomes worse, two mechanisms cause $(\Delta \phi)^2$ to grow. For $\phi=0$ and $C_2 = 1$, the imperfect contrast $C_1$ leads to a smaller signal with measurement noise corresponding to only $N C_1$ atoms, so $\left(\Delta \phi(0)\right)^2=1/(N C_1)$ in this case. However, when $C_2 < 1$, all $N$ atoms contribute to the $S_z$ measurement noise, giving $\left(\Delta \phi(0)\right)^2=1/(N C_2^2)$. As $|\phi|$ increases, $\Delta \phi$ becomes equal for both situations because the $N (1 - C_1)$ atoms decohered during state preparation are rotated toward the equator and contribute their projection noise to the final $S_z$ measurement.
The corresponding clock stabilities are shown in Figure \ref{fig:PhaseErrorAndClockStabilityWithContrastLoss}(b) and (d), where we see that the best Ramsey time remains near $\gamma \tau \approx 0.5$ and the optimum clock performance scales as $\sigma^2_\phi \propto C_1^{-1}$ and $\sigma^2_\phi \propto C_2^{-2}$, just like $\left(\Delta \phi(0)\right)^2$.

\begin{figure}[hbtp]
	\centering
	\includegraphics[width = 1\columnwidth]{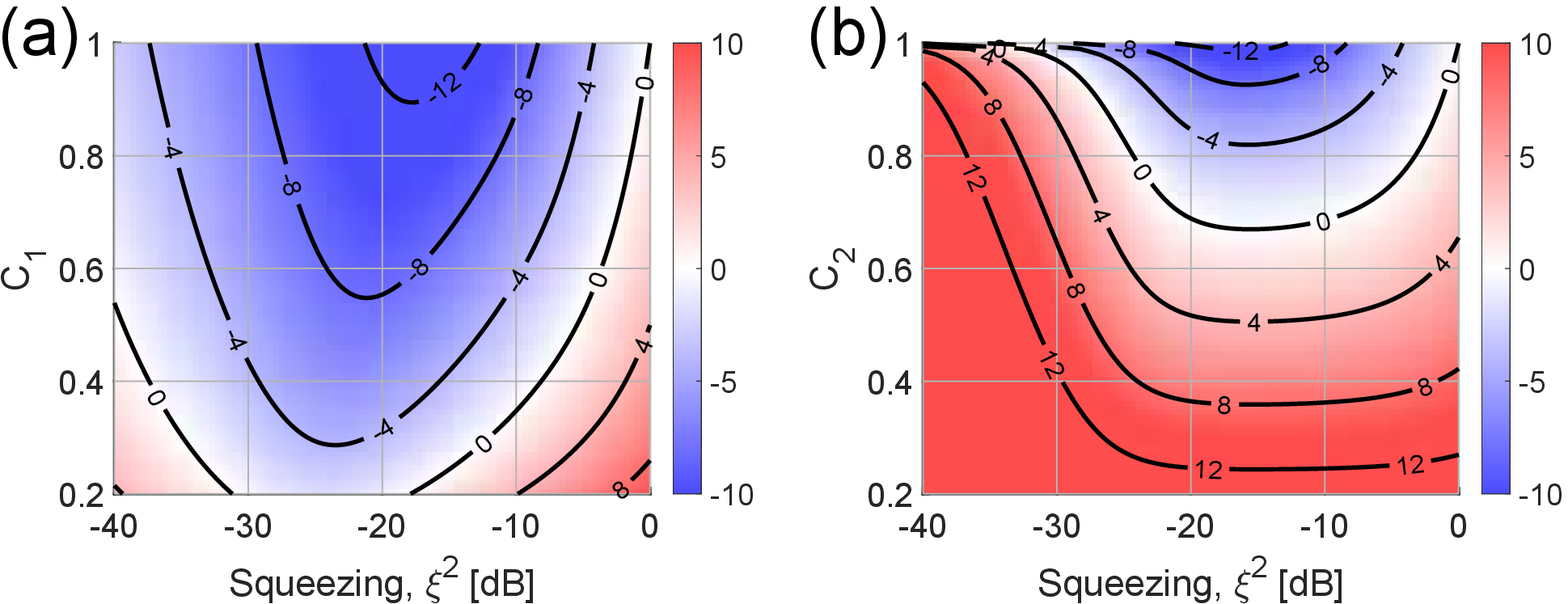}
	\caption{
		Plot of the best possible clock performance, in units of $\mathrm{dB}$ of clock phase variance, compared to a clock with no squeezing and perfect contrast for $N=10^4$ atoms, as a function of the amount of squeezing $\xi$ and (a) preparation contrast $C_1$ and (b) Ramsey time contrast $C_2$. We assume no excess antisqueezing, $\xi \chi = 1$, the usage of measurement squeezing ($\theta = 0$), and unity fixed contrast $C_2=1$ and $C_1=1$ in (a) and (b) respectively.
	}  
	\label{fig:BestClockStabilityVsSqueezingAndContrast}
\end{figure}

Finally, we focus on the potential gain in clock stability, optimizing over the Ramsey time $\tau$. Figure \ref{fig:BestClockStabilityVsSqueezingAndContrast} shows the effect of contrast loss on the performance of a squeezed clock with $N=10^4$ atoms. The colors and contours are in units of $\mathrm{dB}$, normalized to the SQL: a clock with no squeezing and perfect contrast (top-right corner of these plots). As expected, contrast loss $C_2$ during the Ramsey time has a much more severe effect on ultimate clock stability than imperfections in the initial state $C_1$. We can see that if the former contrast falls below $C_2=0.7$, the clock performance will always be worse than a clock without squeezing and unity contrast. More interestingly, as $C_2$ becomes smaller, the potential improvement from squeezing also decreases -- for $C_2=1$, we can gain as much as $-13.7\un{dB}$ in clock stability by squeezing, but for $C_2=0.4$, this decreases to only $-1.7 \un{dB}$. This finding is consistent with the results of \cite{Huelga1997,Ulam-Orgikh2001}: when single-particle loss limits the Ramsey time, squeezing is unable to significantly improve clock stability.

\section*{References}

\bibliography{UnitarySqueezingAndClocks}
\bibliographystyle{unsrt}

\end{document}